\documentclass[aps,twocolumn,showpacs,superscriptaddress,amsmath,amssymb]{revtex4}
\usepackage{graphicx}
\usepackage{dcolumn}
\usepackage{bm}
\begin{document}
\preprint{NUHEP-EXP/07-03}
\title{Detecting Neutrino Magnetic Moments with Conducting Loops}
\author{Aram Apyan}
\email{aapyan@iit.edu}
\affiliation{Illinois Institute of Technology,
Physics Division, 3101 South Dearborn St., Chicago, IL. 60616, USA}
\author{Armen Apyan}
\email{aapyan@lotus.phys.northwestern.edu}
\author{Michael Schmitt}
\email{schmittm@lotus.phys.northwestern.edu}
\affiliation{Northwestern University, Department of Physics and Astronomy, 2145
Sheridan Road, Evanston, IL 60208, USA}
\newcommand{\EMF}       {{\sc EMF}}
\newcommand{\emfmax}    {\varepsilon_{\mathrm{max}}} 
\newcommand{\tmax}      {t_{\mathrm{max}}} 
\newcommand{\Wtot}      {W_{\mathrm{tot}}}
\date{September 23, 2007}

\begin{abstract}
It is well established that neutrinos have mass, yet it is very difficult
to measure those masses directly.  Within the standard model of particle
physics, neutrinos will have an intrinsic magnetic moment proportional to
their mass.  We examine the possibility of detecting the magnetic moment
using a conducting loop.  According to Faraday's Law of Induction, 
a magnetic dipole passing through a conducting loop induces an 
electromotive force, or \EMF, in the loop.  
We compute this \EMF\ for neutrinos in several cases,
based on a fully covariant formulation of the problem.  We discuss
prospects for a real experiment, as well as the possibility to test
the relativistic formulation of intrinsic magnetic moments.
\end{abstract}

\pacs{41.20.Gz, 71.15.Rf, 87.50.Mn, 83.60.Np}

\maketitle

\section{\label{sec:intro}Introduction}
\par
Neutrinos are known to have very small masses, though it is not known yet 
what those masses are~\cite{degouvea}.  It appears to be difficult to probe 
masses much below the~eV-level with current techniques, so new techniques
are desirable.  We would like to take advantage of the {\em linear}
relation between the mass of an elementary particle and its magnetic 
dipole moment, and find a way to measure directly the magnetic moment of 
neutrinos. Our suggestion is to exploit Faraday's Law of Induction 
which states that a change in magnetic flux through a conducting loop
induces a current in that loop.  Neutrinos in the lab are entirely
left-handed, so the passage of a beam of neutrinos through a conducting
loop will result in pulses in the loop of alternating direction.
The purpose of this paper is to estimate the basic size and shape
of such pulses.
\par
We consider a conceptual experiment consisting of a small conducting ring
through which a neutral elementary particle passes with a known trajectory 
and momentum.  If that particle possesses an intrinsic
magnetic dipole moment arising from its spin, then the passage of the particle
through the loop induces an electromotive force (\EMF) due to the change
of magnetic flux through the surface bounded by the ring.
The \EMF\ is proportional to the magnetic dipole moment.  
This method was successfully applied for measuring the magnetic 
moments of neutrons and various nuclei~\cite{bloch1,bloch2} and in
searches for magnetic monopoles~\cite{tassie,eberhard,soma}.
\par
Faraday's Law of Induction has been treated in classical electrodynamics 
mainly for non-relativistic velocities of magnets and rings. 
Cullwick~\cite{cullwick} is one of the most valuable references that 
addresses Faraday's Law for speeds comparable to the speed of light. 
\par
We shall calculate the induced \EMF\ for different alignments of the ring
with respect to the particle momentum. We will see that an analytic solution 
is possible only for the case when particle passes through the center of 
the ring, and its direction of motion is perpendicular to the plane of the ring.
For other configurations we provide numerical solutions.

\section{\label{mag:mmt}Lorentz Transformations of the Magnetic Dipole Moment}
\par
If a particle at rest has an intrinsic electric dipole moment  
$ \boldsymbol d $, and it is placed in an external electric field  
$ \boldsymbol E $, its potential energy is  $ \boldsymbol {- d \cdot E} $. 
Similarly, a particle with a magnetic dipole moment  $ \boldsymbol \mu $ has a 
potential energy  $ \boldsymbol {-\mu \cdot B} $ when it is placed in an external 
magnetic field  $ \boldsymbol B $. We need to express these terms in
covariant form, using the anti-symmetric $4$-tensor for the electric
and magnetic fields,~$F_{\mu\nu}$.  The appropriate covariant generalization
of the electric and magnetic three-vectors $({\boldsymbol{d}},{\boldsymbol{\mu}})$
is another anti-symmetric tensor, defined by
\begin{equation}
\sigma^{\mu \nu} = \left ( \begin{array}{cccc} 0 & c d_x & c d_y & c d_z \\
-c d_x & 0 & -\mu_z & \mu_y \\ -c d_y & \mu_z & 0 & -\mu_x \\ -c d_z & -\mu_y & \mu_x & 0 \end{array} \right)
\label{eq:matrix}
\end{equation}
where $c$ is the speed of light in vacuum.  The famous ``BMT Equation''~\cite{BMT}
is consistent with this formulation.  On this basis, the ``interaction energy'' is 
written~\cite{barut,vanHolten}
\begin{equation}
U = \boldsymbol {- d \cdot E} - \boldsymbol {\mu \cdot B} = \frac {1} {2} 
\sigma^{\mu \nu} F_{\mu \nu}
\label{eq:mdm-def}
\end{equation}
which indicates that $U$ is a Lorentz scalar.
\par
The transformation equations for the electric and magnetic dipole 
moments under a Lorentz boost are now easily obtained:
\begin{eqnarray}
\boldsymbol {\mu} &=& \gamma \boldsymbol {\mu'} + (1-\gamma) \frac { \boldsymbol {v \cdot \mu'}} {v^2} \boldsymbol v +  \gamma  \boldsymbol {v \times
d'}  \nonumber \\
\boldsymbol {d} &=& \gamma \boldsymbol {d'} + (1-\gamma) \frac { \boldsymbol {v \cdot d'}} {v^2} \boldsymbol v - \frac {\gamma} {c^2} \boldsymbol {v \times
\mu'} 
\label{eq:transform}
\end{eqnarray}
where $\gamma$ is the Lorentz factor and $\boldsymbol v $ is the relative 
velocity of the coordinate frames. If we assume that the electric dipole 
moment is zero, and the velocity $ \boldsymbol v $ has an $x$-component only, then 
from Eq.~(\ref{eq:transform}) we have
\begin{eqnarray}
\mu_x&=& \mu_x' \nonumber \\
\mu_y&=& \gamma \mu_y' \nonumber \\
\mu_z&=& \gamma \mu_z'
\label{eq:transform1}  
\end{eqnarray}
where the primed quantities refer to the rest frame of the particle.

\section{\label{mag:emf}The Induced EMF in the Ring}
\par
\begin{figure}[htbp]
\begin{center}
\includegraphics[scale=0.42]{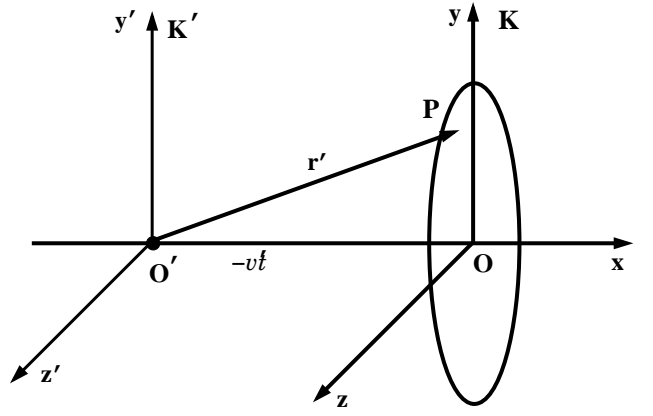}
\end{center}
\caption{coordinate frames $K$ and $K^{\prime}$, associated with the ring and
particle, respectively}
\label{F:frame}
\end{figure}

We introduce two coordinate systems~(see Fig.~\ref{F:frame}) to calculate 
the induced \EMF\ when the particle passes through the center of a ring
of radius~$b$, and its motion is perpendicular to the plane of the ring.
At $t = t' = 0$ the origin of the laboratory coordinate system~($K$),
attached to the ring, coincides with the origin of the coordinate system~($K'$)
attached to the particle. We will need to express $r^\prime$ in terms of the 
coordinates in $K$.  From the Lorentz transformations we have
\begin{eqnarray}
x' &=& \gamma(x - v t) \nonumber \\
r' &=& \sqrt{ \gamma^2 ( x - v t )^2 +{y}^2 + {z}^2 } .
\label{eq:radius}
\end{eqnarray}

\vfill

\par
The magnetic field of the dipole in $K'$ is  
\begin{equation}
\mathbf{B'} (\mathbf{r'}) = \frac{\mu_0} {4 \pi} 
\frac{1} {{r'}^3} \bigg [
     3 (\boldsymbol{\mu'} \cdot \mathbf{\hat{r}'}) \mathbf{\hat{r}'} - 
         \\\boldsymbol {\mu'} \bigg ]
\label{eq:magfield}
\end{equation}
where $\mu_0$ is the permeability of the vacuum, $\boldsymbol \mu'$  is the magnetic
dipole moment, $\mathbf r'$ is a radius vector in $K'$, and $\mathbf{\hat{r}'}$ 
is the unit vector in the direction of $\mathbf r'$. By expressing each 
quantity on the right-hand side of this equation by its respective quantity in 
$K$ (for points on the plane of the ring $x=0$), and then performing Lorentz 
field transformations ($ \boldsymbol E' = 0$), we obtain the electromagnetic 
field-tensor components in the laboratory frame:
\begin{widetext}
\begin{eqnarray}
E_x & =& 0,  \nonumber \\
E_y & =& \frac{ 3 \mu_0 v} {4 \pi (\gamma^2v^2t^2+y^2+z^2)^{5/2}} 
\Big [- \mu_x \gamma^2 v t z +\mu_y y z + \mu_z z^2 - \frac {\mu_z} {3} 
\big(\gamma^2v^2t^2+y^2+z^2 \big) \Big] ,  \nonumber \\
E_z &=& \frac{ - 3 \mu_0 v} {4 \pi (\gamma^2v^2t^2 +y^2 +z^2)^{5/2}} 
\Big [-\mu_x \gamma^2 v t y + \mu_y y^2 + \mu_z y z - \frac {\mu_y} {3}
\big(\gamma^2 v^2 t^2 +y^2 + z^2\big)\Big], \nonumber \\
B_x &=& \frac{ 3 \mu_0 } {4 \pi (\gamma^2v^2t^2+y^2+z^2)^{5/2} }
\Big[\mu_x \gamma^2 v^2 t^2  - \mu_y v t y - \mu_z v t z - \frac {\mu_x} {3} 
\big(\gamma^2 v^2 t^2 +y^2 + z^2\big)\Big], \nonumber \\
B_y &=& \frac{ 3 \mu_0 } {4 \pi (\gamma^2v^2t^2 +y^2 +z^2)^{5/2}} 
\Big[-\mu_x \gamma^2 v t y + \mu_y y^2 + \mu_z y z - \frac {\mu_y} {3} 
\big(\gamma^2 v^2 t^2 +y^2 + z^2\big)\Big], \nonumber \\
B_z & =& \frac{ 3 \mu_0 } {4 \pi (\gamma^2v^2t^2+y^2+z^2)^{5/2}} 
\Big[- \mu_x \gamma^2 v t z +\mu_y y z + \mu_z z^2 - \frac {\mu_z} {3}
\big(\gamma^2v^2t^2+y^2+z^2\big)\Big] 
\label{eq:field}
\end{eqnarray}
\end{widetext}

\par
We assume that the particle moves uniformly through the ring and its energy 
loss is negligible. When the particle passes through the ring there is 
an induced current in the ring. Thus, there will be a force (see 
Eq.~\ref{eq:mdm-def}) on the particle as it approaches the ring
\begin{equation}
\boldsymbol F = \boldsymbol { \nabla(\mu \cdot  B) } = \mu_x \frac 
{\partial B_x} {dx} \mathbf {\hat x}  .
\end{equation}
The work done on the particle by 
this force is negligible compared to the energy of the particle even for a 
strong constant current in the ring ($W = -\mu_0\mu_x I/b$). 
When the particle passes through the center, the induced current reverses its 
direction, and therefore the particle continues to dissipate energy, 
consistent with the conservation of energy. 

\par
To calculate the induced \EMF, we parameterize the path of the integration in 
counterclockwise direction from positive $x$~\cite{apostol}
\begin{equation}
\varepsilon = \oint_{c} \mathbf{E} \, \cdot \, {d}\boldsymbol{\ell}
 = \int_{0}^{2\pi} \mathbf {E} (\boldsymbol {\rho} (\theta)) \cdot \frac {\mathbf {d} \boldsymbol {\rho}} {d \theta} d\theta    
\label{eq:emf}
\end{equation}
where $\boldsymbol \rho (\theta) = b \cos\theta \mathbf {\hat y} + b 
\sin\theta \mathbf {\hat z }$ is the parameterization of the curve for
parameter $\theta \in [0,2\pi]$. We have the following geometric relations
\begin{eqnarray}
\frac {d\boldsymbol{\rho(\theta)}}{d\theta} &=& 
-b \sin\theta\, \mathbf{\hat y} + b\cos\theta\, \mathbf{\hat z } \nonumber \\
y^{2} + z^{2} &=& b^{2} .
\end{eqnarray}
Integration yields
\begin{equation}
\varepsilon = \frac{3} {2}  \mu_0 \mu_x b^2  \frac{ \gamma^2 v^2 t} 
{(\gamma^2 v^2 t^2 + b^2)^{5 / 2} } ,
\label{eq:no16}
\end{equation}
and for relativistic cases,
\begin{equation}
\varepsilon \approx \frac{3} {2}  \mu_0 \mu_x b^2  \frac{ c^2 \gamma^2 t}
{ \big [c^2 \gamma^2 t^2 + b^2 \big ]^{5 / 2} }
\label{eq:no18}
\end{equation}

\par
We can calculate the induced \EMF\ by a second method.  Define the flux
$\varphi = \int \mathbf {B} \cdot \mathbf {d} \boldsymbol {\sigma} $ 
and calculate the time derivative ${d\varphi}/{dt}$,
\begin{equation}
\frac{d \varphi}{dt} = 
  \int_{S}^{} \frac{\partial\mathbf{B}}{\partial t} 
    \cdot {d}\boldsymbol {\sigma} = 
   \int_{0}^{2 \pi} \int_{0}^{b} 
   \frac {\partial \mathbf {B}} {\partial t} \cdot ( \frac {\partial \mathbf {r}} 
 {\partial \rho} \times \frac {\partial \mathbf {r}} {\partial \theta} ) 
  \, d\rho \, d\theta
\end{equation}
where  $ \mathbf r (\rho,\theta) = \rho \cos\theta \mathbf {\hat y} 
+ \rho \sin\theta \mathbf {\hat z } $ is the parameterization of the surface; 
$ \theta \in [0,2\pi] $ and $  \rho \in [0,b] $ are the angular and
radial parameters.
From the right-hand rule and the direction of $d\boldsymbol{\ell}$,
it follows that $d\boldsymbol{\sigma}$ (the element of 
surface area of ring) should point in $\hat{\boldsymbol x}$ direction. 
Indeed, $(\frac {\partial \boldsymbol {r}} {\partial \rho} \times \frac 
{\partial \mathbf {r}} {\partial \theta} )= \rho \mathbf {\hat x }$.
Therefore,
\begin{equation}
\frac{d\varphi} {dt} = \int_{0}^{2 \pi} \int_{0}^{b} \frac {\partial {B_x}} {\partial t}
 \, \rho \, d\rho \, d \theta
\end{equation}
and integration yields
\begin{equation}
\frac{d\varphi} {dt} = -\frac{3} {2}  \mu_0 \mu_x b^2  \frac{ \gamma^2 v^2 t} 
{(\gamma^2 v^2 t^2 + b^2)^{5 / 2} }
\label{eq:dphidt}
\end{equation}
which is consistent with Eq.~(\ref{eq:no16}).

\par
\begin{figure}[htbp]
\begin{center}
\includegraphics[scale=0.43]{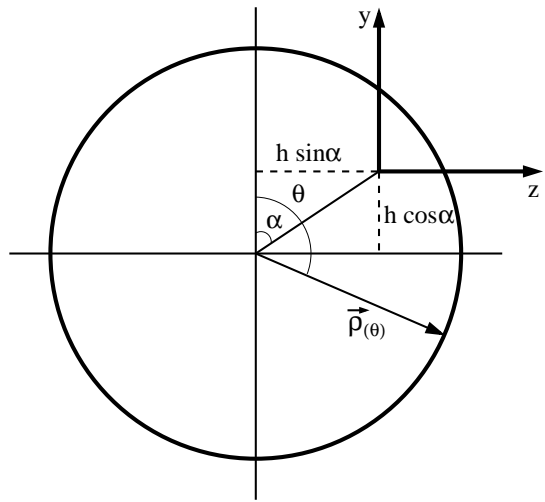}
\end{center}
\caption{coordinates for off-axis passage, $ (h, \alpha ) $
\label{F:circle}}
\end{figure}

To calculate the induced \EMF\ when the particle does not pass through the 
center of the ring, we need to modify the parameterization of the curve along 
which the line integral (Eq.~\ref{eq:emf}) is calculated.  
We will continue to assume that the particle trajectory is perpendicular to 
the plane of the ring. The parametrization of the curve describing the
conducting loop is now (Fig.~\ref{F:circle})
\begin{equation}
 \boldsymbol \rho (\theta) = (b \cos\theta - h\cos \alpha ) \mathbf {\hat y} + 
(b \sin\theta - h \sin \alpha ) \mathbf {\hat z} 
\label{eq:par}
\end{equation}
From symmetry it follows that the induced \EMF\ is independent of $\alpha$, 
so we set $\alpha = 0$.  We have  the following geometric relations:
\begin{eqnarray}
  \frac{{d}\boldsymbol{\rho(\theta)}} {d \theta} &=& 
  -b \sin\theta\, \mathbf{\hat y} + b \cos\theta\, \mathbf {\hat z } \nonumber \\
y^{2} + z^{2} &=& b^{2} + h^2 - 2bh\cos\theta
\label{eq:geometry}
\end{eqnarray}

\par
The induced \EMF\ can be computed by inserting
Eqs.~(\ref{eq:par}) and~(\ref{eq:geometry}) into 
Eq.~(\ref{eq:emf}). There is no analytical solution, 
however, so the integral has been evaluated numerically.
\par
In the general case, the particle moves in arbitrary direction toward the ring. 
The velocity $\boldsymbol v$ defines the $x$-axis.  We define the orientation
of the ring by  two angles $\phi$ and $\beta$. Let's define the coordinate system 
$ (\ell,m,n) $, where $m$ and  $n$ axes define the plane of the ring and
$ \ell $ is directed along the normal to the plane. The origins of the 
$ (\ell,m,n) $ and $ (x,y,z) $ coordinate systems coincide. Performing two
rotations in three dimensions (by angle $\phi$ relative to $\ell$ and by angle
$\beta$ relative to $x$), we can relate the coordinates of a point in 
$ (\ell,m,n) $  to the  $ (x,y,z) $ coordinate system. We can define the $m$-axis
to be along the line of the nodes (the intersection line of the $yz$ plane and 
the plane of the ring)~\cite{giacovazzo}, so that one of the Eulerian angles 
is zero. The other two Eulerian angles have simple geometric meanings, namely,
$\phi$ is the angle between the $x$-axis and the normal of the ring 
($\ell$ axis), and $ \beta $ is the angle between the $y$-axis and the 
line of nodes ($m$-axis). We can parameterize $ \ell $, $m$, $n$ similarly as in 
Section~\ref{mag:emf}:
\begin{widetext}
\begin{eqnarray}
\boldsymbol{\rho}(\theta) &=&  -\sin\phi (b \sin\theta - h\sin\alpha ) 
\mathbf {\hat x} + [\cos\beta (\cos\theta - h\cos\alpha) + \cos\phi \sin\beta 
(b\sin\theta - h\sin\alpha)] \mathbf {\hat y} + 
{} 
\nonumber \\ & & {} 
[\cos \beta \cos\phi (b\sin\theta - h \sin\alpha) - \sin\beta 
(b\cos\theta- h \cos\beta)] 
\mathbf {\hat z}
\end{eqnarray}
\end{widetext}

\section{\label{mag:current}The Induced Current in the Ring}
\par
To find the current in the ring, one must solve
the following differential equation
\begin{equation}
 \frac{d\phi}{dt} + L\frac{dI}{dt} + IR = 0
\label{eq:circuit}
\end{equation}
where $R$ is the resistance of the ring, $L$ is its inductance, and
\begin{equation}
  \varepsilon = -\left[ \frac{d\phi}{dt} + L\frac{dI}{dt} \right] .
\label{eq:emfcircuit}
\end{equation}
For $R \neq 0$, the second term in Eq.~(\ref{eq:circuit}) is
negligible compared to the third, and one finds trivially that
the current is simply the \EMF\ divided by the resistance.
\par
For superconductors, $R = 0$, and it follows from Eq.~(\ref{eq:emfcircuit})
that the \EMF\ is zero.  From Eq.~(\ref{eq:circuit}) we obtain
$d(\phi+LI)/dt = 0$, which states that the magnetic flux through
a superconducting loop is constant in time.  One says that the
magnetic field lines are ``frozen'' in the ring, a result which
can also be obtained directly from the field equations.  We have
already calculated $d\phi/dt$ -- see Eq.~(\ref{eq:dphidt}).
Solving this differential equation we have
\begin{equation}
  I(t) = - \frac{\mu_0\mu_x b^2}{2L}
           \frac{1}{(c^2\gamma^2 t^2 + b^2)^{3/2}}
\label{eq:current}
\end{equation}
where the initial current is taken to be zero.  The negative sign
follows from the orientation of the ring that we chose in calculating
Eq.~(\ref{eq:dphidt}).

\section{\label{mag:rd}Results and Discussion}
\par
In the simple case in which the particle passes normally through the center
of the ring, the \EMF\ does not depend on the transverse components 
$\mu_y$ and $\mu_z$ of the magnetic dipole moment.  
The only contribution to the \EMF\ comes from $\mu_x$, the component of magnetic  
moment in the direction of motion. 

One should investigate Eq.~(\ref{eq:no18}) for the maximum and minimum
values of the \EMF. By taking the derivative of the equation over time we have
\begin{equation}
\emfmax = \pm \frac{3} {4} \bigg(\frac{4} {5} \bigg)^{5/2} \mu_0 
\mu_x c \, \frac{\gamma} {b^2}
\label{eq:max-emf}
\end{equation}
corresponding to the times
\begin{equation}
\tmax = \pm \frac {b} {2 c {\gamma}} 
\label{eq:time-emf}
\end{equation}
which depend on the geometry of the configuration and the energy of the particle.

\begin{figure}[htbp]
\begin{center}
\includegraphics[scale=0.4]{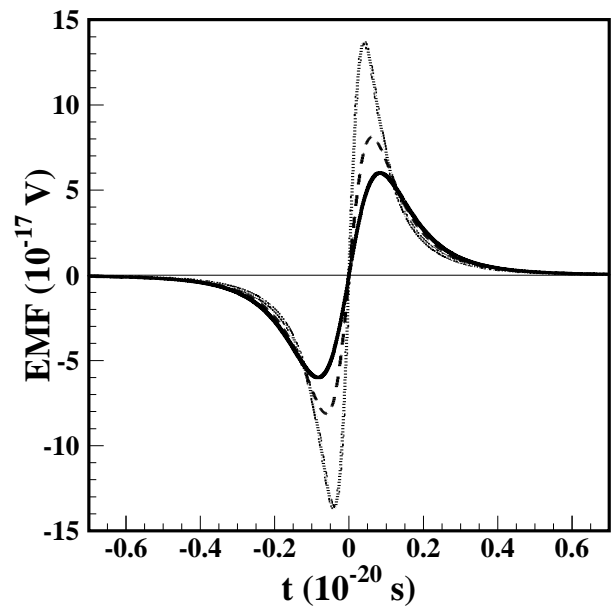}
\end{center}
\caption{induced \EMF\  in the ring when particle passes through the ring 
on-axis~(solid line) and off-axis (dashed line for $h = 0.4 b$ and dotted 
line for $ h = 0.6 b$)}
\label{F:on-axis}
\end{figure}

We present  the induced \EMF\ in the ring as a function of time
in Fig.~\ref{F:on-axis}. For this example, we took an electron neutrino~($\nu_e$) 
with energy $10$~GeV and a ring with diameter of $1$~cm. We assume that neutrino 
has a mass of $M_\nu = 1$~eV and the magnetic moment is $\mu_x = 10^{-10} \mu_B$, 
where $\mu_B$ is the Bohr magneton. This mass and magnetic moment are close to
the current upper limits~\cite{pdg}.  The two extremum values of the \EMF\ are 
seen to be in agreement with Eq.~(\ref{eq:max-emf}) and~(\ref{eq:time-emf}).
The solid curve represents the induced \EMF\ when the particle passes through the center 
and is perpendicular to the surface of the ring. The dashed and dotted curves 
represent the induced \EMF\ when the particle passes through the ring with distances 
from its center of $0.4 b$ and $0.6 b$, respectively. 

\par
As seen in Fig.~\ref{F:on-axis}, the induced \EMF\ from a single relativistic
neutrino is extremely small, on the order of $10^{-17}$~V, and the pulse
is extremely brief, on the order of $10^{-21}$~s. For off-axis trajectories, 
the effective radius of the ring appears smaller and as a result the induced 
\EMF\ is larger (dashed and dotted curves in Fig.~\ref{F:on-axis})
and occurs at shorter times~\cite{soma}. 
\par
\begin{figure}[htbp]
\begin{center}
\includegraphics[scale=0.4]{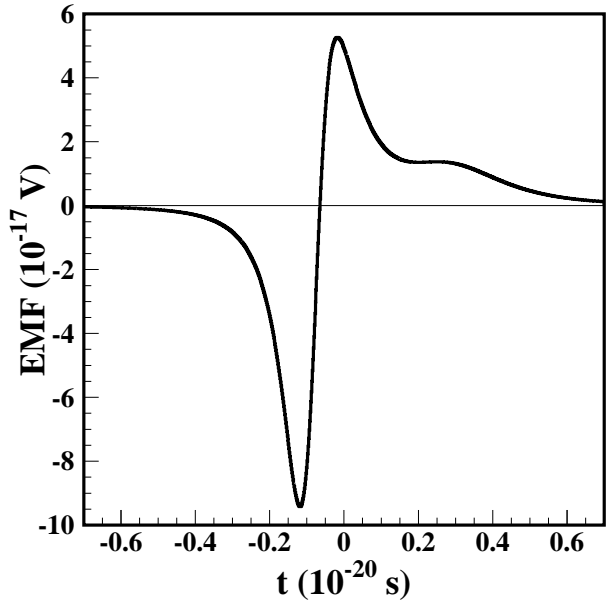}
\end{center}
\caption{induced \EMF\ when the particle trajectory makes an angle 
$\phi = 0.1$~nrad to the normal of the ring}
\label{F:asym}
\end{figure}

An example for the \EMF\ in the general case is presented in Fig.~\ref{F:asym},
for which $h = 0.6b$, $\phi = 0.1$~nrad, $\beta = 0$, and  $\alpha = \pi/2$. 
Due to the tilt, there is an asymmetry between the maximum 
and minimum \EMF\ which might be useful in an experiment.
However, in this case the \EMF\ decreases very rapidly with $\phi$. 

\par
Could a practical detector be built to observe this signal?
Certainly special techniques would be required to detect
such a small and brief pulse.  It is not easy to increase the
signal, since increasing~$\gamma$ or decreasing~$b$ will 
increase~$\emfmax$ but decrease~$\tmax$; and in fact,
$\int_0^\infty \varepsilon(t)\, dt = 1.5 \mu_0 \mu_x / b$,
independent of~$\gamma$.   If the loop has resistance~$R$, 
then the power dissipated in the loop for all time is
\begin{equation}
 \Wtot = K\, \bigg(\frac{3}{2}\mu_0\bigg)^2 \,\frac{\gamma v \mu_x^2}{b^3R}
\label{eq:power}
\end{equation}
where $K = \int_{-\infty}^{\infty} dx\, x^2(1+x^2)^{-5} = 5\pi/128$.
If the magnetic moment is proportional to the mass,
$\mu_x = A M_\nu$, as in the standard model~\cite{Fujikawa},
then $\Wtot \approx A^2 E_\nu M_\nu / b^3R$,
which suggests that $M_\nu$ might be measured from this technique.
However, for our on-axis example depicted in Fig.~\ref{F:on-axis},
and for $R = 1~\Omega$,  $\Wtot \approx 10^{-53}$~W, which is
exceedingly small.
\par
The case of a superconducting ring is also interesting.  The current $I(t)$
was presented in Eq.~(\ref{eq:current}).  It is very interesting that the
current does not change direction as a function of time.  Its maximum magnitude,
however, is extremely small, on the order of $10^{-45}$~A.
\par
Neutrinos produced at an accelerator are bunched in time, so
one might try to increase the signal from the coherent contributions 
of a large number of neutrinos ($N_\nu = {\cal{O}}(10^6)$), effectively 
boosting~$\mu_x$ by that number ({\rm cf.} Eq.~(\ref{eq:max-emf}) and~(\ref{eq:power})).
However, the neutrinos would have to arrive at the loop with time
differences small compared to~$\tmax$, and since the bunching of 
neutrinos occurs on the scale of $10^{-12}$~s, which is very long 
compared to~$\tmax \approx 250~\mathrm{fm}/c$, this seems impractical.
\par
Apart from any possibility of measuring the magnetic moment
of the neutrino, we would like to mention the opportunity
of investigating experimentally the transformation of the
intrinsic spin of a particle under Lorentz boosts.  
Eq.~(\ref{eq:no16}) predicts a dramatic variation of the
\EMF\ as a function of the particle's velocity, and
Eq.~(\ref{eq:power}) indicates that $\Wtot \propto \beta\gamma$,
which rises linearly with $v \ll c$ and then increases much
more rapidly for $v > 0.6 c$.   The neutron has a magnetic
moment of $-1.91\mu_N$, where $\mu_N$ is the nuclear magneton,
which is roughly $2000$~times smaller than~$\mu_B$.  
The neutron's magnetic moment is therefore a factor $10^7$~times
larger than the moments we have considered for the neutrino.
However, for neutron energies $E_n = 10$~GeV, $\gamma \approx 10$,
which is much less than the factor $10^{10}$ assumed for the
neutrino, so the signal is again quite small:
$\emfmax = 7\times 10^{-19}$~V, $\tmax = 0.8$~ps,
and $\Wtot = 10^{-48}$~W. For non-relativistic neutrons,
the signal would be even smaller.  Nonetheless, a 
confirmation of Eq.~(\ref{eq:no16}) or Eq.~(\ref{eq:max-emf})
would imply that the Ansatz for the relativistic
transformation of intrinsic spin, as given in  
Eqs.~(\ref{eq:matrix}) and~(\ref{eq:mdm-def}), is correct.

\begin{acknowledgments}
We are grateful to Prof. A. Garg for helpful discussions.
\end{acknowledgments}
\bibliography{induction3.bib}

\begin{thebibliography}{14}
\expandafter\ifx\csname natexlab\endcsname\relax\def\natexlab#1{#1}\fi
\expandafter\ifx\csname bibnamefont\endcsname\relax
  \def\bibnamefont#1{#1}\fi
\expandafter\ifx\csname bibfnamefont\endcsname\relax
  \def\bibfnamefont#1{#1}\fi
\expandafter\ifx\csname citenamefont\endcsname\relax
  \def\citenamefont#1{#1}\fi
\expandafter\ifx\csname url\endcsname\relax
  \def\url#1{\texttt{#1}}\fi
\expandafter\ifx\csname urlprefix\endcsname\relax\def\urlprefix{URL }\fi
\providecommand{\bibinfo}[2]{#2}
\providecommand{\eprint}[2][]{\url{#2}}

\bibitem[{\citenamefont{de~Gouv\^ea}(2004)}]{degouvea}
\bibinfo{author}{\bibfnamefont{A.}~\bibnamefont{de~Gouv\^ea}},
  \bibinfo{journal}{Mod. Phys. Lett.} \textbf{\bibinfo{volume}{A19}},
  \bibinfo{pages}{2799} (\bibinfo{year}{2004}).

\bibitem[{\citenamefont{Alvarez and Bloch}(1940)}]{bloch1}
\bibinfo{author}{\bibfnamefont{L.~W.} \bibnamefont{Alvarez}} \bibnamefont{and}
  \bibinfo{author}{\bibfnamefont{F.}~\bibnamefont{Bloch}},
  \bibinfo{journal}{Phys.\ Rev.} \textbf{\bibinfo{volume}{57}},
  \bibinfo{pages}{111} (\bibinfo{year}{1940}).

\bibitem[{\citenamefont{Bloch}(1946)}]{bloch2}
\bibinfo{author}{\bibfnamefont{F.}~\bibnamefont{Bloch}},
  \bibinfo{journal}{Phys.\ Rev.} \textbf{\bibinfo{volume}{70}},
  \bibinfo{pages}{460} (\bibinfo{year}{1946}).

\bibitem[{\citenamefont{Tassie}(1965)}]{tassie}
\bibinfo{author}{\bibfnamefont{L.~J.} \bibnamefont{Tassie}},
  \bibinfo{journal}{Nuovo Cimento} \textbf{\bibinfo{volume}{38}},
  \bibinfo{pages}{1935} (\bibinfo{year}{1965}).

\bibitem[{\citenamefont{Eberhard et~al.}(1971)\citenamefont{Eberhard, Ross, and
  Alvarez}}]{eberhard}
\bibinfo{author}{\bibfnamefont{P.~H.} \bibnamefont{Eberhard}},
  \bibinfo{author}{\bibfnamefont{R.~R.} \bibnamefont{Ross}}, \bibnamefont{and}
  \bibinfo{author}{\bibfnamefont{L.~W.} \bibnamefont{Alvarez}},
  \bibinfo{journal}{Phys.\ Rev} \textbf{\bibinfo{volume}{4}},
  \bibinfo{pages}{3260} (\bibinfo{year}{1971}).

\bibitem[{\citenamefont{Somalwar et~al.}(1988)\citenamefont{Somalwar, Frisch,
  and Incandela}}]{soma}
\bibinfo{author}{\bibfnamefont{S.}~\bibnamefont{Somalwar}},
  \bibinfo{author}{\bibfnamefont{H.}~\bibnamefont{Frisch}}, \bibnamefont{and}
  \bibinfo{author}{\bibfnamefont{J.}~\bibnamefont{Incandela}},
  \bibinfo{journal}{Phys.\ Rev D} \textbf{\bibinfo{volume}{37}},
  \bibinfo{pages}{2403} (\bibinfo{year}{1988}).

\bibitem[{\citenamefont{Cullwick}(1957)}]{cullwick}
\bibinfo{author}{\bibfnamefont{E.~G.} \bibnamefont{Cullwick}},
  \emph{\bibinfo{title}{Electromagnetism and Relativity, with particular
  reference to moving media and electromagnetic induction}}
  (\bibinfo{publisher}{Longmans, Green and Co}, \bibinfo{address}{London},
  \bibinfo{year}{1957}).

\bibitem[{\citenamefont{Bargmann et~al.}(1959)\citenamefont{Bargmann, Michel,
  and Telegdi}}]{BMT}
\bibinfo{author}{\bibfnamefont{V.}~\bibnamefont{Bargmann}},
  \bibinfo{author}{\bibfnamefont{L.}~\bibnamefont{Michel}}, \bibnamefont{and}
  \bibinfo{author}{\bibfnamefont{V.~L.} \bibnamefont{Telegdi}},
  \bibinfo{journal}{Phys. Rev. Lett.} \textbf{\bibinfo{volume}{2}},
  \bibinfo{pages}{435} (\bibinfo{year}{1959}).

\bibitem[{\citenamefont{Barut}(1980)}]{barut}
\bibinfo{author}{\bibfnamefont{A.~O.} \bibnamefont{Barut}},
  \emph{\bibinfo{title}{Electrodynamics and Classical Theory of Fields and
  Particles}} (\bibinfo{publisher}{Dover}, \bibinfo{address}{New York},
  \bibinfo{year}{1980}).

\bibitem[{\citenamefont{van Holten}(1992)}]{vanHolten}
\bibinfo{author}{\bibfnamefont{J.~W.} \bibnamefont{van Holten}}
  (\bibinfo{year}{1992}), \eprint{hep-th/9303124}.

\bibitem[{\citenamefont{Apostol}(1969)}]{apostol}
\bibinfo{author}{\bibfnamefont{T.~M.} \bibnamefont{Apostol}},
  \emph{\bibinfo{title}{Calculus v.2}} (\bibinfo{publisher}{John Wiley \&
  Sons}, \bibinfo{address}{New York}, \bibinfo{year}{1969}).

\bibitem[{\citenamefont{Giacovazzo}(2002)}]{giacovazzo}
\bibinfo{author}{\bibfnamefont{E.~C.} \bibnamefont{Giacovazzo}},
  \emph{\bibinfo{title}{Fundamentals of Crystallography}}
  (\bibinfo{publisher}{Oxford University Press}, \bibinfo{address}{New York},
  \bibinfo{year}{2002}).

\bibitem[{\citenamefont{Yao and others [PDG~Collaboration]}(2006)}]{pdg}
\bibinfo{author}{\bibfnamefont{W.-M.} \bibnamefont{Yao}} \bibnamefont{and}
  \bibinfo{author}{\bibnamefont{others [PDG~Collaboration]}},
  \bibinfo{journal}{Journal of Physics} \textbf{\bibinfo{volume}{G33}},
  \bibinfo{pages}{1} (\bibinfo{year}{2006}).

\bibitem[{\citenamefont{Fujikawa and Shrock}(1980)}]{Fujikawa}
\bibinfo{author}{\bibfnamefont{K.}~\bibnamefont{Fujikawa}} \bibnamefont{and}
  \bibinfo{author}{\bibfnamefont{R.}~\bibnamefont{Shrock}},
  \bibinfo{journal}{Phys. Rev. Lett.} \textbf{\bibinfo{volume}{45}},
  \bibinfo{pages}{963} (\bibinfo{year}{1980}).

\end{thebibliography}
\end{document}